# About Digital Communication Methods for Visible Light Communication

Wataru Uemura, Yasuhiro Fukumori and Takato Hayama

Department of Electronics and Informatics, Ryukoku University, Shiga, Japan

## Abstract

*The visible light communication (VLC) by LED is one of the important communication methods because LED can work as high speed and VLC sends the information by high flushing LED. We use the pulse wave modulation for the VLC with LED because LED can be controlled easily by the microcontroller, which has the digital output pins. At the pulse wave modulation, deciding the high and low voltage by the middle voltage when the receiving signal level is amplified is equal to deciding it by the threshold voltage without amplification. In this paper, we proposed two methods that adjust the threshold value using counting the slot number and measuring the signal level. The number of signal slots is constant per one symbol when we use Pulse Position Modulation (PPM). If the number of received signal slots per one symbol time is less than the theoretical value, that means the threshold value is higher than the optimal value. If it is more than the theoretical value, that means the threshold value is lower. So, we can adjust the threshold value using the number of received signal slots. At the second proposed method, the average received signal level is not equal to the signal level because there is a ratio between the number of high slots and low slots. So, we can calculate the threshold value from the average received signal level and the slot ratio. Unfortunately, the first proposed method adjusts the threshold value after receiving the data, once the distance between the sender and the receiver is changed, then the performance becomes worse. And after adjusting the threshold, the performance becomes better. Therefore, this method should be used in stable environments. The second proposed method can change the threshold value during the signal is received. That means this method can work very quickly. So, this method can show good performance for the wide range. We show these performances as real experiments.*

## Keywords

*Visible Light Communication, Digital Communication, Pulse Position Modulation.*

## 1. Introduction

Light Emitting Diode (LED) becomes widespread as a lighting source which can save power consumption. And because the response speed of LED is high, LED is focused on as the sender element of the visible light communication (VLC) which transmits the information by blinking the light [1 - 4]. For visible light communication, pulse modulation is used often because it is easy for the output pin of a microcomputer to control the power of the LED. If we assign 0 and 1 information to the voltages of LED which turn off and on (which are high and low voltage, respectively), flicker has occurred because the lighting time (that means turning on time) is changed by the information. One of modulation methods to avoid the flicker is Pulse Position Modulation (PPM). On PPM, the lighting time per information is constant [5, 6]. When the distance between the sender and the receiver becomes larger, there are more communication errors like the other wireless communications because of the decreasing receiving signal level [7]. The gain control for the receiving signal is important for this problem. If we use the pulse modulation, the decision method which compares the amplified voltage as the half voltage of high is equal to the decision method which compares the received voltage as the threshold voltage. In this paper, we propose two methods for the receiving signal level; one method adjusts the threshold voltage, and the other method controls the gain for the received voltage automatically. The number of signal slots is constant per one symbol when we use PPM. If the number of received signal slot per one symbol time is less than the theoretical value, that means the threshold value is higher than the optimal value. If it is more than the theoretical value, that means the threshold value is lower. So, we can adjust the threshold value using the number of received signal slot. At the second proposed method, the average received signal level is not equal to the high signal level because there is the ratio between the number of high slots and the number of low slots. So, we can calculate the threshold value from the average received signal level and the slot ratio.

Section 2 shows the communication method which uses visible light instead of the radio wave. And Section 3 describes the comparator voltage which uses to decode the signal. We proposed two methods in Section 4. And we show the experimental results in Section 5. Finally, we conclude this paper in Section 6.

## 2. THE VISIBLE LIGHT COMMUNICATION

VLC is wireless communication that uses light instead of the radio wave [8 - 10]. In this section, we describe the modulation method for visible light communication and its applications.

### 2.1. The Frequency and Modulation of the Visible Light Communication

VLC is one of the wireless communications. It uses light for communication, which is visible to the human eyes and is one of the electromagnetic waves like the infrared. We can see the visible light which frequency range is about from 360-400 [nm] to 760-830 [nm]. We cannot sense the electromagnetic waves which frequency is over about 830 [nm]. The shorter wavelengths than the visible light is called ultraviolet rays and shorter than it is called as X rays and gamma rays. And the longer one is called infrared rays and longer than it is called the radio wave.

At VLC, we send the information by the changing blinking patterns of the light. The LED is an element for the lighting source of visible light communication. LED has some better features than traditional lighting elements, for example, the low power consumption, the long lifetime, the highspeed response, and the small size. From the view of the highspeed response, LED can turn on and off very quickly. And by this feature, VLC can communicate at a very high speed. And from this quick response, we can have a variety of modulations for VLC. In this paper, we use LED as the lighting source.

### 2.2. The Features of VLC

In this subsection, we show some characteristics of visible light communication when we use it as wireless communication. At first, we can see the communication area because the medium of the communication is visible light. This characteristic is only for visible light communication comparing with other wireless communications. And it is easy to cut and change the communication range by using a wall, a mirror, a lens, and so on. In such a case, we can modify the communication range by checking it with our eyes. By the reason of using the ordinal lighting elements, it is not affected to the human, that means it is safe for the body, and in addition to it, it is not affected to the other devices using the radio wave. Furthermore, it has high directivity, so we can control the directional of communication and we can get a high resolution. But there is a defect, for which there are a lot of noise sources which are natural and artificial lighting. In the remainder of this section, we describe the features of visible light communication comparing with other electromagnetic wave communication, like radio and infrared.

## 2.3. Comparison with the Radio Wave Communication

The radio wave communication is the wireless communication with the electromagnetic wave which frequency is less than 3 [THz] under the Japanese radio law. The radio wave has some features; the refraction, the diffraction, and reflection regarding its frequency. And the wavelength becomes shorter, directivity becomes stronger.

It is difficult for us to find the source of the radio wave because it has more refraction than visible light. So, it is difficult for the sender to control the area where the information can be transmitted. That means some security issues will happen more frequently than VLC. And there is a possibility that the radio wave affects human health. Besides, because it affects electric devices, it is prohibited to use radio wave communication in a hospital, a train, an airplane, and so on. And the radio law limits the power and the frequency used by radio wave communication.

On the other hand, VLC has the following features against radio wave communication. Because the visible light goes straight without few refractions, the obstructs can cut off the communication. It is easy for the receiver to find the sender location because the transmitter is bright. And it is easy for the sender not to send the information to the undesirable receivers because the sender can control the receivable range. If the power of the visible light is too big, it is dangerous for us same as the radio wave. But if we face such a situation, we can avoid it because we recognize its dangerously by eyes. And we can use VLC in a hospital, a train, an airplane, and so on because of no effect on the electrical device. Also, VLC is not under the radio law, so we can set any power and any frequency for VLC as you want.

## 2.4. Comparison with the Infrared Communication

The infrared communication is the wireless communication with the infrared wave which frequency is longer than red light and shorter than the radio waves, and its wavelength is about from 0.7 [μm] to 1 [mm]. Its characteristics from the view of the electronic wave and visible are similar to that of visible light because the frequency of the infrared is close to that of the visible light.

However, the infrared wave cannot be seen by a human, so the transmitter of the infrared cannot be lighting that we cannot use as visible elements. And then the human cannot recognize the communication range. It is dangerous for us to use more than certain strengths because we cannot see the effect for the human eyes. Furthermore, we cannot find where the transmitter device is set on and cannot see the communication range. Then there is the possibility that the receiver whom the sender does not want to send the information to can receiver gets the packet, and if that will happen, the sender cannot know it. This is not good for security, so we need to use infrared communication to avoid this incident.

From the view of VLC, it has similar characteristics against infrared communication because the frequency of it is very close to the frequency of infrared. But some characteristics of VLC are different for infrared. One of the most different characteristics is visibility.

## 2.5. About Modulations

We need to modulate the signal to send the information using wireless communications. The modulation is to change the frequency, amplitude, or phase of the carrier to send the information by the signal. Usually, in digital communication, the electronic signal is changed by the information '0' and '1'.

There are a lot of kinds of modulation. For example, we can change the strength of LED or the pattern of LED for visible light communication. We call the amplitude modulation (AM) where we change the voltage of LED regarding the sending information. And we call the frequency modulation (FM) where we change the frequency as the LED bringing speed regarding the sending information. And if we switch the power of LED to on or off regarding the information, we call it as on-off keying (OOK) [11]. If we change the duty rate of the pulse regarding the sending information, it is called the pulse width modulation (PWM) [12]. If we change pulse position in the time slots regarding the sending information, it is called pulse position modulation (PPM). When we use the modulation, which changes the something of pulse, usually the pulse forms not the sine curve but the rectangular. For the considering of the usage for the lighting equivalent, we need to not occur the flicker and need to brightness light [13, 14]. In this paper, we use the PPM as the lighting modulation method because it keeps the same lighting power for any information.

PPM change the time slot of the symbol which is one of the sending information. One symbol allows only one pulse which width is one slot. So, this modulation has only a single lighting slot per one symbol for any information. Then the flicker has not occurred because the average power of a certain time is the same as any information. If the number of positions per symbol becomes larger, we can send more information per one symbol. For example, we use 2-PPM for 2 slots per symbol, 4-PPM for 4 slots per symbol (shown in Figures 1 and 2, respectively). Furthermore, we can send the information even if the state of the slot is changed to upside down, for example, the high becomes low and low becomes high. We call it Inversed-PPM (shown in Figure 3). If we use 2-PPM, 2-IPPM is the same as 2-PPM. When we use more than two -PPM, like 4-PPM, the brightness is different between PPM and IPPM. For example, using 4-IPPM, the lighting time ratio of on and off is 3:1, the brightness can be suppressed to 3/4. So, the lighting time becomes longer, and the average power becomes bigger. Then from the human eyes, the lighting LED of the transmitter becomes bright. This characteristic is good for lighting equipment.

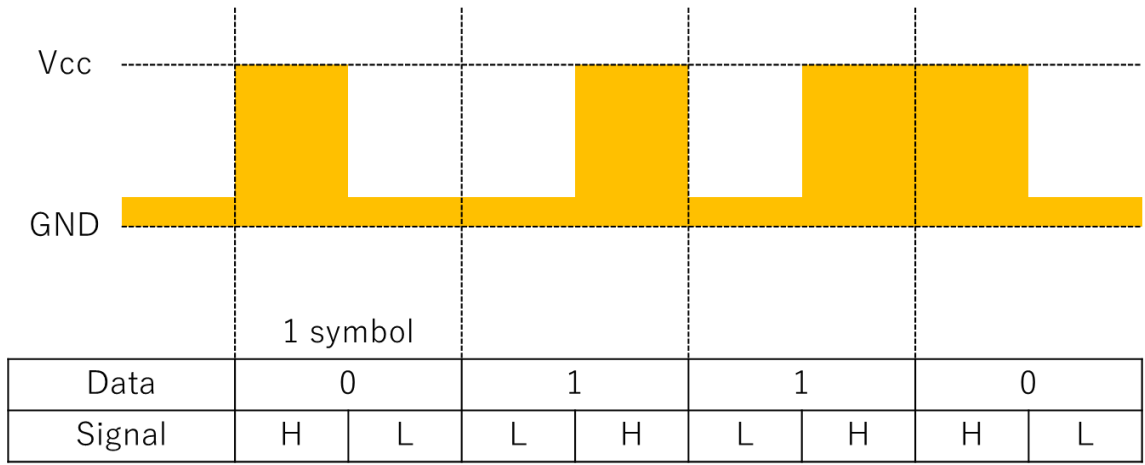

Figure 1. 2-PPM for 2 slots per symbol

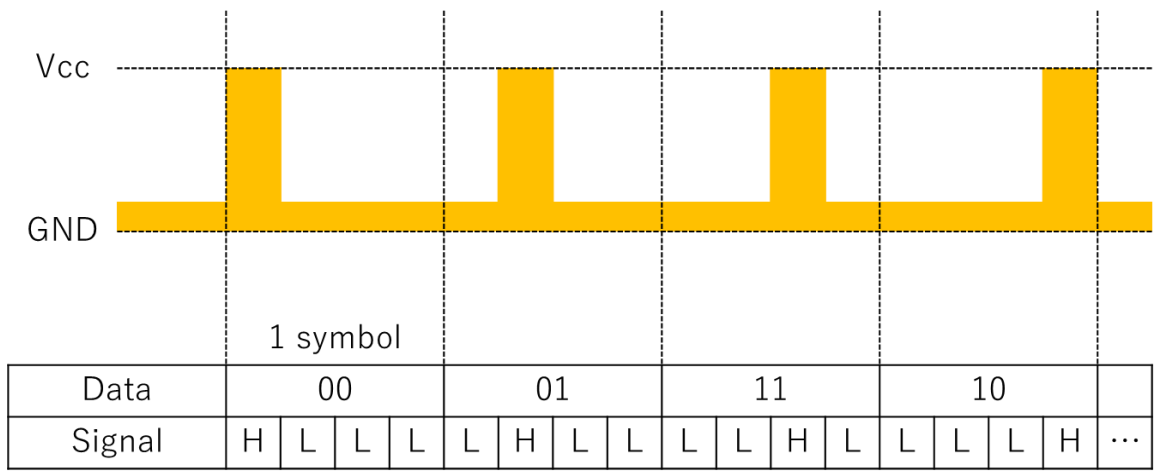

Figure 2. 4-PPM for 4 slots per symbol

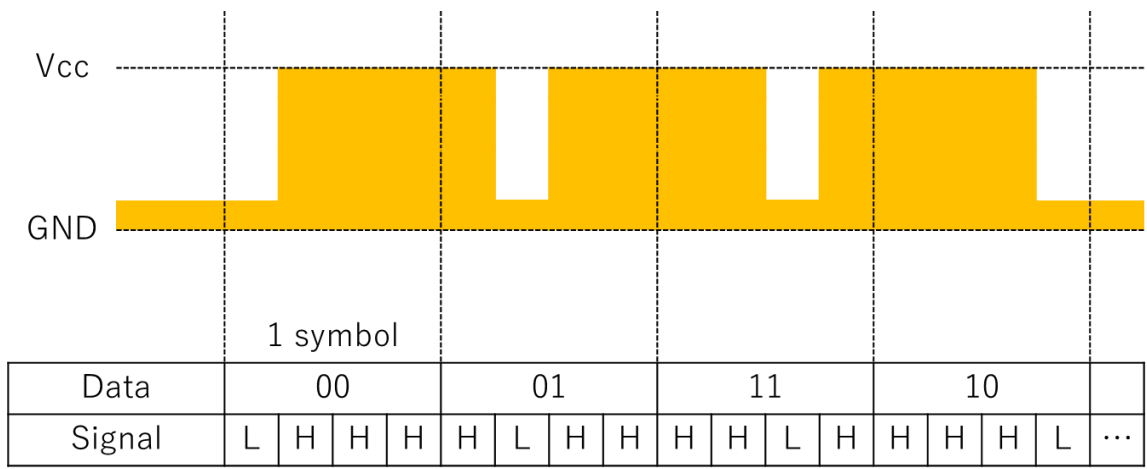

Figure 3. 4-Inversed PPM for 4 slots per symbol

Generally, a filter circuit is used for reducing the influence of noise. However, when a filter circuit is used in visible light communication with a rectangular wave, its waveform is deformed, and the edge of the rectangular wave is lost. And it becomes difficult to decide 0 or 1 with its deformed waveform. Now we focus on the comparator circuit which compares two input voltages and outputs high or low voltages according to its result. By setting the threshold value as fixing the input voltage, if the input voltage exceeds a threshold voltage, the output becomes high voltage, otherwise, the output becomes low voltage. As a result, the output waves become the rectangular wave that does not include an unnecessary waveform other than the transmission waveform, and the influence of noise can be reduced. Here, in the visible light communication using the rectangular wave, there is no problem even if the received signal is deformed into a rectangular wave shape by the comparator circuit. Therefore, by using the threshold voltage set by the comparator as a reference, the waveform can be adjusted by the comparator, and communication with fewer errors can be performed. For example, waveforms deformed by the filter circuit, waveforms distorted by a noise that cannot be completely cut, and waveforms distorted by noise generated in the circuit, can be adjusted.

## 3. ADJUSTING THE COMPARATOR VOLTAGE

The receiver sets threshold voltage for determining high and low from the signal voltage received by the lighting element. Here, as in other communication methods, the influence of noise is a problem to decode the signal. In this paper, we propose a comparative voltage adjustment method.

### 3.1. About Noise in Digital Communication of Visible Light

A filter is used for reducing the influence of noise. There are several filters; a low-pass filter that cuts high-frequency signal, a high-pass filter that cuts low-frequency signal, a bandpass filter that cuts frequencies other than certain frequencies, and so on. In this way, filters cut noise from specific frequencies. However, the noise like white noise generally does not depend on the frequency and affects all frequency bands, so that noise remains even after being filtered. On the other hand, when rectangular waves are used as modulation, the shape of the wave is known, and the waveform can be adjusted by the comparator. By preparing a threshold value to distinguish between high and low from the signal, we can create square waves by outputting using only high and low voltages.

### 3.2. Problems in Setting the Reference Voltage

Usually, the threshold value for the comparator is manually adjusted using a variable resistor. In this case, if the communication distance between the sender and the receiver becomes longer, the signal strength decreases, and it becomes impossible to distinguish between high and low by the fixed threshold value. Therefore, the threshold value adjusted first is not always appropriate. We need to adjust the threshold value of the comparator.

### 3.3. About Adjusting the Threshold Value for the Comparator

In this subsection, we discuss the setting method for the comparator threshold value.

The best value for the threshold at the comparator should be half value of the signal voltages. But it is depended on the distance between the sender and the receiver because the receiving signal levels are decreasing as long as its distance. Considering this effect, it is the good value which is a little bigger value than the low voltage, for example, +0.1 [v]. If we set such a value as the threshold value, it has the effectiveness to decode the signal for the long distance.

Unfortunately, in the real world, there is noise that changes the signal level from low to over the threshold. One of the methods which avoid this noise, we can use the comparator element like in the previous Section. The comparator outputs the high voltage if the input level is more than the threshold voltage. Otherwise, it outputs the low voltage. By using the comparator, we can output the just rectangle wave which includes only the signal when the threshold value can be set more than the noise voltage. However, the receiving signal level will be changed whenever the distance between the sender and the receiver is changed. After that, some parts of the output waveform will be changed to high or low form according to its receiving signal level, and it causes the bit errors. Figure 4 shows the result by three kinds of threshold values. Usually, we adjust the threshold value manually according to the communication distance.

### 3.4. Threshold Detection Methods

If there is no noise signal, we should set the threshold value to a little bigger value than 0 [v]. And if we get the noise information, we can calculate the threshold value using the noise voltage. M. Usman et al. discussed the characteristics of photo receivers and light sources. And they estimated the noise distribution [15]. The goodness of the fit is determined by Bayesian Information Criterion and the non-centrality parameter of the corresponding Rician distribution is used to calculate the decision threshold.

On the other hand, Y. H. Kim et al. used the optical and color filters to decide the threshold [16]. They estimate the noise level using the differential value between the received signal and the threshold value is be changed regarding the differential.

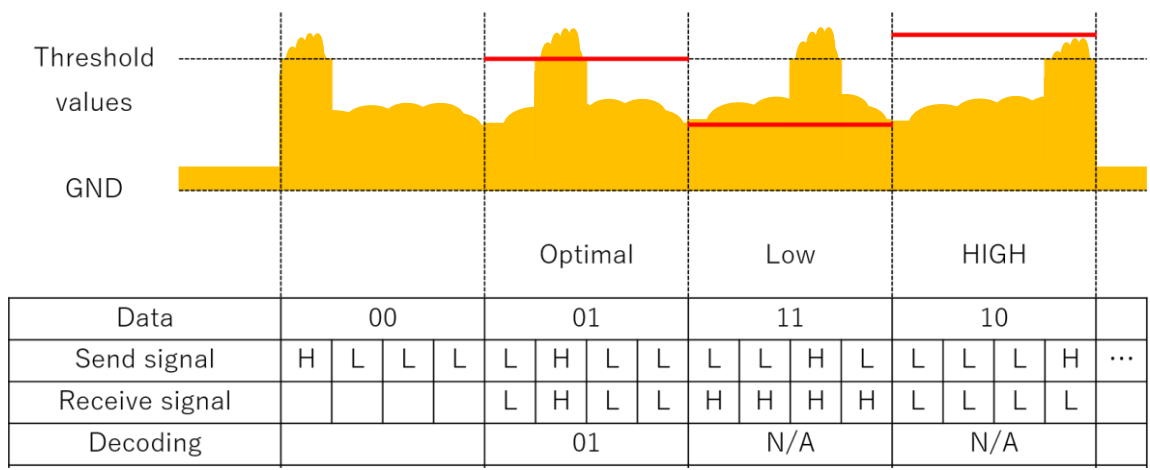

Figure 4. This figure shows how to adjust the threshold value. If there is no noise, the optimal value for the threshold value is a little more than 0V. But due to noise, if we use such value for the threshold, there are a lot of high signal. And if we set the threshold value more than the optimal value, we do not receive the signal. We adjust the threshold value using the variable register.

## 4. AUTOMATIC ADJUSTMENT OF THRESHOLD VALUE OF THE COMPARATOR

We propose the method for adjusting a comparison voltage of a comparator according to a change in voltage of a received signal. In this Section we propose the following two adjusting methods. Both methods use the PPM. In the first method, we focus on the number of signal slots. The number of PPM signal slots is constant per one symbol. If the number of received signal slot per one symbol time is less than the theoretical value, that means the threshold value is higher than the optimal value. If it is more than the theoretical value, that means the threshold value is lower. So, we can adjust the threshold value using the number of received signal slot. In the second method, we focus on the average received signal level. The average received signal level is not equal to the high signal level. Because there is the ratio between the number of high's slots and the number of low's slots. So, we can calculate the threshold value from the average received signal level and the slot ratio.

## 4.1. Setting the Threshold Value from the Receiving Slot Information

At first, we propose the adjusting method which uses the slot rate for high and low. When we use the constant value for the threshold to decide the receiving signal is high or low, if the receiving signal is stronger, the high slot number is more than the low slot number in the output slots. And if the receiving signal is weaker, the low slot number is more than the high. The slot ratio of high and low is 1:1 at the 2-PPM, so we change the threshold value larger or smaller until the ratio becomes 1:1. When we use 4-PPM, the ratio is 1:3. Figure 5 which includes the noise signal, shows the relationship between the threshold value and the received signal level. If the number of received signal slot per one symbol time is more than the theoretical value, that means the threshold value is lower than the optimal value. If it is less than the theoretical value, that means the threshold value is higher. So, we can adjust the threshold value using the number of received signal slot. Figure 6 shows a block diagram of its automatic adjustment circuit.

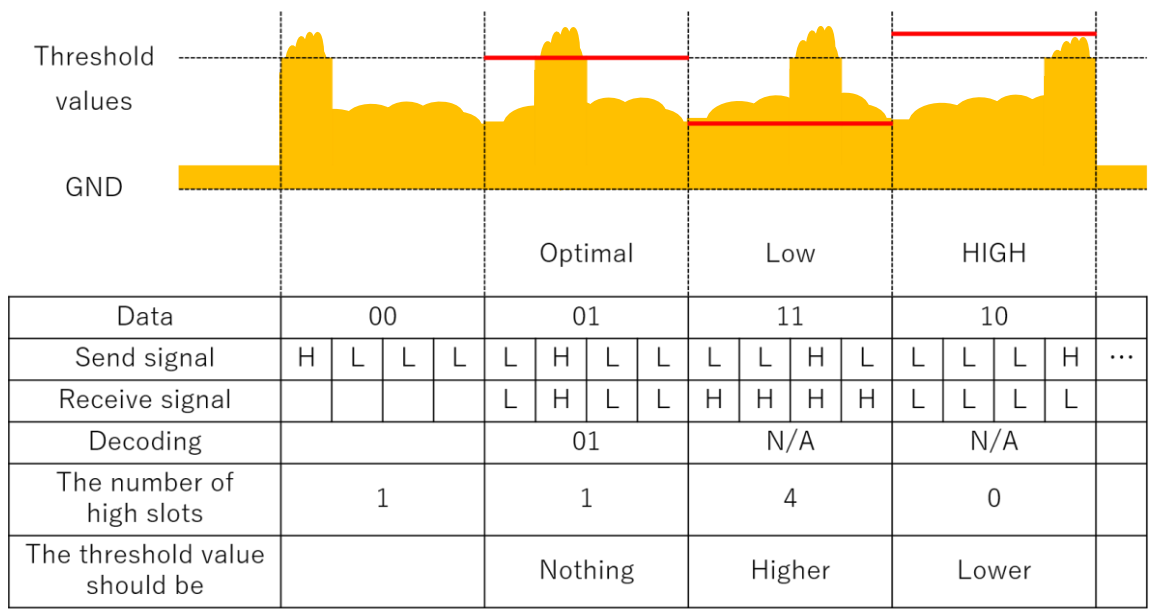

Figure 5. This figure shows how to adjust the threshold value. If the threshold value is less than the optimal value, the number of high slots is more than 1. Then the threshold value should become higher. If the threshold value is more than the optimal value, the number of high slots is less than 1 (= 0). Then the threshold value should become lower.

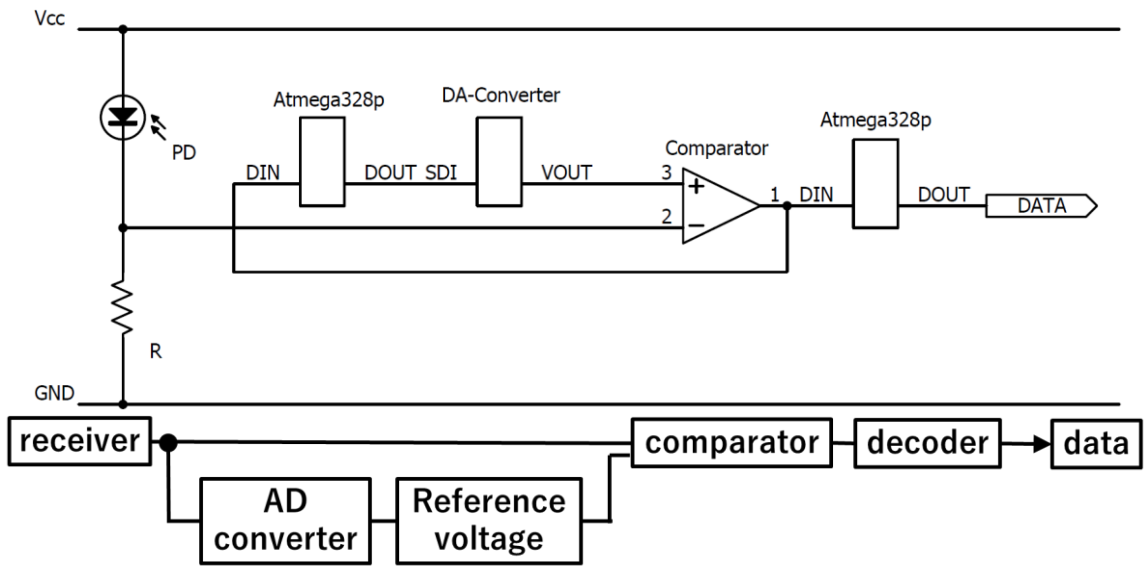

Figure 6. The circuit and block diagram of proposed method which change the reference value step by step in order to adjust the ratio of receiving high and low slots.

## 4.2. Set the Threshold Value from the Receiving Voltage

Second, we propose the adjusting method which uses the threshold value from the receiving voltage. Figure 7 shows a block diagram of its automatic adjustment circuit. To determine the threshold value, it is important to obtain the voltage value of the signal, but the analogue converter in the microcomputer operates slowly. Therefore, an analogue-digital (AD) converter IC is used. From the view of the receiver, the voltage of the high signal and the voltage of the low signal (expressed as $V_A$[V] and $V_B$[V], respectively) fluctuates depending on the distance between the transmitter and the receiver. So, they must be decided from the signal level of the receiving waveform. Then, the average voltage of the received signal $V_{average}$ is $(V_A^2 + V_B^2) / (T_A + T_B)$, where $T_A$ and $T_B$ are the signal time, respectively. When $T_A : T_B = 1 : 1$, then $V_{average} = (V_A + V_B) / 2$, and this is suitable for the threshold value.

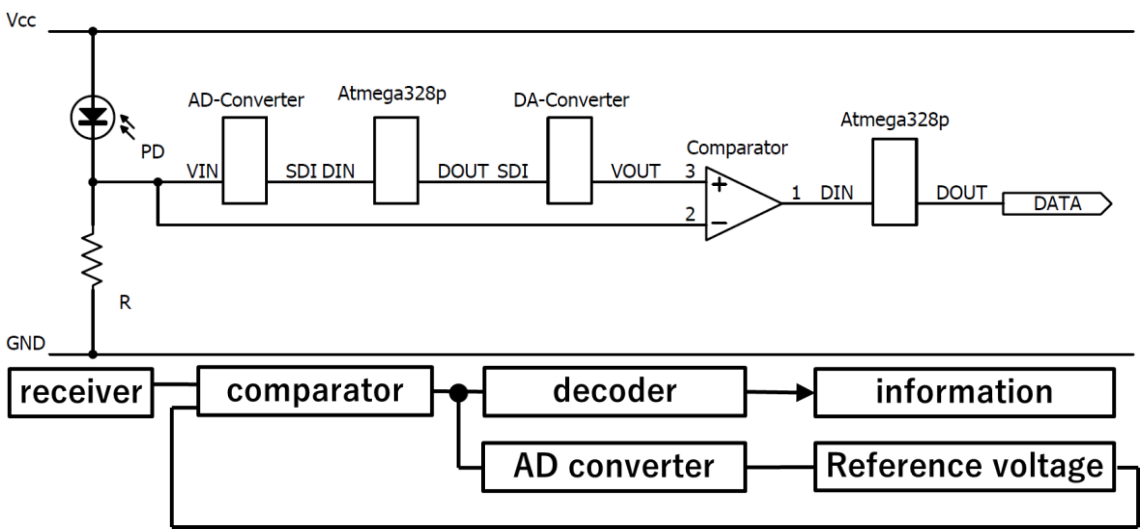

Figure 7. The circuit and block diagram of proposed method which changes the reference value step by step in order to adjust the receiving signal level

## 5. EXPERIMENTAL RESULTS

### 5.1. About the Threshold Value by Counting the Slot Number for High Voltage

For the first proposed method which changes the threshold value by counting the slot number for high voltage, we measure the communication throughput for distances from 1 [cm] to 60 [cm]. We send the pseudo-random binary sequence (PRBS) and calculate the throughput [bps] from the correct data and its receiving time.

Figure 8 shows the performance by the conventional method and our proposed method. The conventional method sets the constant threshold value which adjusts at the 1 cm distance between the sender and the receiver. So over about 23 [cm], the receiver cannot decode the information because the threshold value is over than the received high slot voltage. On the other hand, our proposed method can decode the information for a wide range, but its throughput is not good because our proposed method adjusts the threshold value after receiving data. That means the adjusting time is slow against the receiving data speed. The performance is dependent on the stay time at the same point. So, this method will be good for the slow speed or the stable receiver.

### 5.2. About the Threshold Value by Calculating the Average of Receiving Signal Level

Second, we measure the three methods to show the performance for our second proposed method which adjusts the threshold by the receiving voltage. Three methods are as follows; 1) the constant threshold adjusting for the close distance, 2) the auto-adjusting threshold which is our proposed method, and 3) the constant threshold adjusting for the long distance. We got the bit error rate for the points which distances between the sender and the receiver are from 10 [cm] to 90 [cm]. The communication speed is 4,000 bits per second, and we sent 100,000 bits of information for each point. Figure 9 shows the experimental results by three methods. If we use the constant threshold adjusted for the close distance, we can receive the data to 20 [cm].

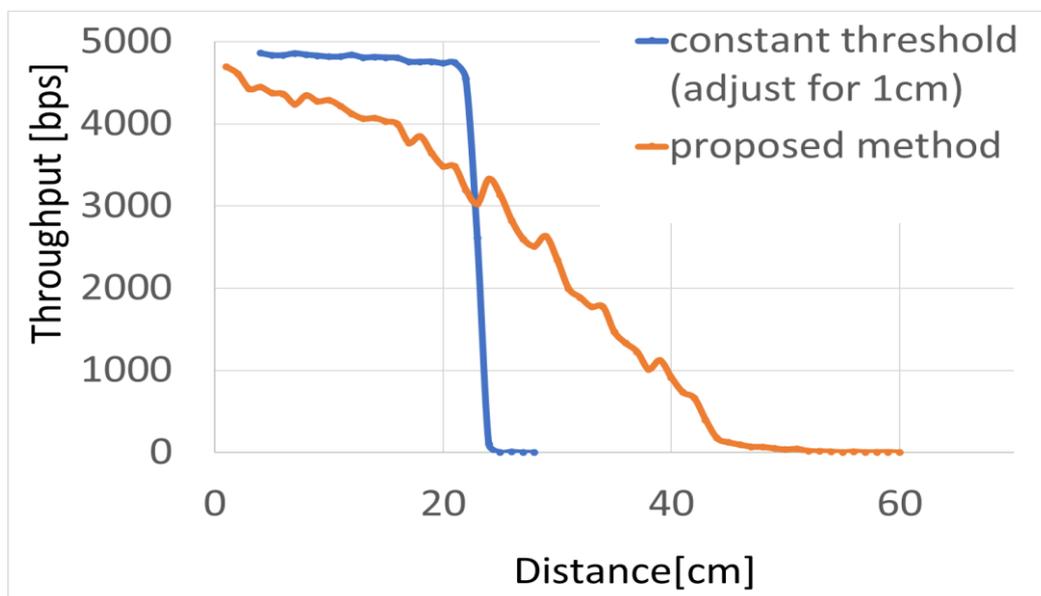

Figure 8. The experimental result shows the communication throughput from 1cm to 60cm.

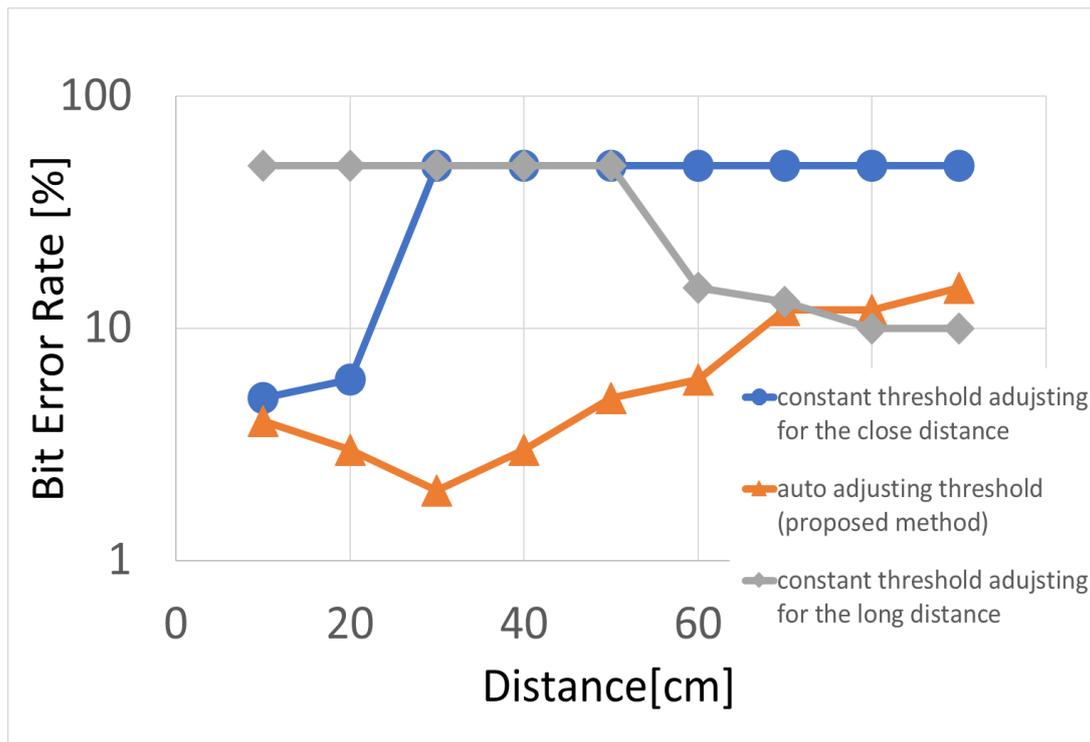

Figure 9. The experimental result shows the communication performances by three methods; the constant thresholds for close and long distance, and proposed method which adjusts the threshold automatically.

But we cannot decode the data because the threshold value is over the received signal level from 20 [cm]. And if we use the constant threshold adjusted for the long-distance, we cannot decode the data to 50 [cm] because the non-signal level with noise is more than the threshold value. Both constant threshold methods are not good performance against the distance which is not assumed by the method. We confirmed the good effect by adjusting the threshold automatically.

## 6. CONCLUSIONS

The visible light communication by LED is focused on because LED can work as high speed and the VLC sends the information by high flushing LED. We use the pulse wave modulation for the VLC with LED because LED can be controlled by the micro controller, which has the digital output pin. At the pulse wave modulation, deciding the high and low voltage by the middle voltage when the receiving signal level is amplified is equal to deciding it by the threshold voltage without amplification. In this paper, we proposed two methods which adjust the threshold value using counting the slot number and measuring the signal level.

In the first method, we can decide the threshold value by counting the receiving slot because the number of high and low slots is constant in the pulse position modulation. Unfortunately, because this method adjusts the threshold value after receiving the data, once the distance between the sender and the receiver is changed, always the performance becomes worse. And after adjusting the threshold, the performance becomes better. Therefore, this method should be used under the stable environments.

Next, we proposed the method which decides the threshold value from the receiving voltage. This method can change the threshold value during the signal is received. That means this method can work very quick. So, this method can show the good performance for the wide range.

For future work, we will focus on the sensitivity of receiving elements because adjusting them will be equal to adjusting the threshold value.

**CONFLICTS OF INTEREST**

The authors declare no conflict of interest.

International Journal of Computer Networks & Communications (IJCNC) Vol.13, No.3, May 2021

**AUTHORS**

**Wataru Uemura** was born in 1977, and received B.E, M.E. and D.E. degrees from Osaka City University, in 2000, 2002, and 2005. He is an associate professor of the Department of Electronics and Informatics, Faculty of Engineering Science, Ryukoku University in Shiga, Japan. He is a member of IEEE, RoboCup and others.

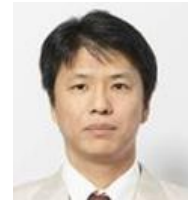

**Yasuhiro Fukumori** was born in 1993 and received B.E, and M.E from Ryukoku University, in 2016, 2018. He is interested in visible light.

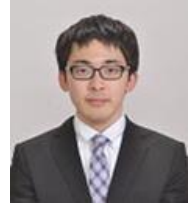

**Takato Hayama** was born in 1992 and received B.E, and M.E from Ryukoku University, in 2015, 2017. He is interested in card games.

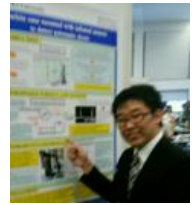